# High-speed ionic synaptic memory based on two-dimensional titanium carbide MXene

Armantas Melianas[1†*], Min-A Kang[2†], Armin VahidMohammadi[3*], Weiqian Tian[2], Yury Gogotsi[3], Alberto Salleo[1*], Mahiar Max Hamedi[2*]

[1] Department of Materials Science and Engineering, Stanford University, Stanford, CA 94305, USA

[2] Department of Fibre and Polymer Technology, School of Engineering Sciences in Chemistry, Biotechnology and Health, KTH Royal Institute of Technology, Teknikringen 56, 10044 Stockholm, Sweden

[3] A. J. Drexel Nanomaterials Institute and Department of Materials Science and Engineering, Drexel University, Philadelphia, PA 19104, USA

† These authors contributed equally

* armantas.melianas@stanford.edu, avm57@drexel.edu, mahiar@kth.se, asalleo@stanford.edu

**Abstract**

Synaptic devices with linear high-speed switching can accelerate learning in artificial neural networks (ANNs) embodied in hardware. Conventional resistive memories however suffer from high write noise and asymmetric conductance tuning, preventing parallel programming of ANN arrays as needed to surpass conventional computing efficiency. Electrochemical random-access memories (ECRAMs), where resistive switching occurs by ion insertion into a redox-active channel address these challenges due to their linear switching and low noise. ECRAMs using two-dimensional (2D) materials and metal oxides suffer from slow ion kinetics, whereas organic ECRAMs enable high-speed operation but face significant challenges towards on-chip integration due to poor temperature stability of polymers. Here, we demonstrate ECRAMs using 2D titanium carbide ($Ti_3C_2T_x$) MXene that combines the high speed of organics and the integration compatibility of inorganic materials in a single high-performance device. Our ECRAMs combine the speed, linearity, write noise, switching energy and endurance metrics essential for parallel acceleration of ANNs, and importantly, they are stable after heat treatment needed for back-end-of-line integration with Si electronics. The high speed and performance of these ECRAMs introduces MXenes, a large family of 2D carbides and nitrides with more than 30 compositions synthesized to date, as very promising candidates for devices operating at the nexus of electrochemistry and electronics.

Inspired by the brain, neuromorphic computing in the form of hardware-based artificial neural networks (ANNs) holds the promise to execute artificial intelligence algorithms at latency and energy efficiency unattainable by conventional computing. To fulfill the potential of hardware ANNs, synaptic devices made with unconventional materials are required because complementary metal-oxide-semiconductor (CMOS) scaling cannot meet the increasingly demanding energy efficiency and computational density requirements of ANNs[1]. For this purpose, two-terminal resistive memories, such as phase-change memory (PCM)[2] and resistive random-access memory (ReRAM)[3], have been investigated owing to their fast switching, small footprint, and integration compatibility. Despite significant progress, these devices are hampered by fundamental limitations such as the programming asymmetry inherent in melting and recrystallization processes in PCMs and the large write errors caused by stochastic filament formation in ReRAMs[4,5]. Furthermore, it has proven difficult to optimize two-terminal resistive memories across all metrics simultaneously[5], such as linear high-speed switching, low write noise, and low current operation, needed for parallel programming of large ANN arrays[6].

To address these challenges, three-terminal synaptic memories, where the write and read operations are decoupled, have recently emerged as a promising alternative. In particular, electrochemical random-access memories (ECRAMs), where resistive switching is facilitated by ion insertion into a redox-active channel, have shown linear switching with low noise in several materials classes, including 2D materials (graphene[7], $WSe_2$[8], α-$MoO_3$[9], α-$Nb_2O_5$[10]), metal oxides ($Li_{1-x}CoO_2$[11], $Li_xTiO_2$[12], $WO_3$[13–15]), and polymers[6,16,17]. Among these, only polymer-ECRAMs have demonstrated the sub-microsecond switching and readout speeds[6,16] compatible with neuromorphic computing. Organic materials however face significant challenges as they do not satisfy the strict temperature and contamination requirements for on-chip integration. 2D material and metal-oxide based ECRAMs on the other hand are more attractive to the semiconductor industry as they are more suitable for on-chip integration[13,14]. These devices are currently severely limited by slow ion kinetics, requiring either millisecond write speeds or millisecond-to-second read pulses to ensure accurate readout of a non-volatile change in device conductance[7–15] (**Supplementary Table S1**). An ideal synaptic memory would combine the high performance and speed of polymer-ECRAMs with the integration compatibility of 2D materials and metal oxides.

**2D MXene enables high-performance ECRAMs**

To find suitable ECRAM materials, one strategy is to investigate emerging redox-active materials, since the ECRAM architecture resembles that of a solid-state electrochemical energy storage device having two thin-film electrodes separated by an electrolyte (**Fig. 1a**). Here, we show that 2D $Ti_3C_2T_x$ MXene enables high-performance and high-speed ECRAMs amenable to future on-chip integration. MXenes are a large family of 2D carbides, nitrides, and carbonitrides with a general formula of $M_{n+1}X_nT_x$, where M is a transition metal (Ti, V, Nb, etc.), X is carbon or nitrogen (or both), n = 1-4, and $T_x$ represents the mixed surface terminations (-O, -OH, -F, -Cl) that are present on the MXene surface[18]. The electronic and electrochemical behavior of MXenes can be tuned by changing their composition, surface terminations, and interlayer chemistry[18,19], resulting in a unique combination of material properties that are highly desirable for ECRAMs. MXenes have been previously explored for threshold-type resistive switches[20–23], but these devices lack non-volatility and face the same challenges as conventional 2-terminal synaptic memories, namely poor linearity and high write noise (**Supplementary Table S1**). Here, we show how these challenges can be overcome by using MXenes as the redox-active resistive switching layer in ECRAMs (**Fig. 1a**). More broadly, we demonstrate the potential of this diverse materials family to enable devices that leverage ion intercalation to switch their electronic properties with unprecedented speed, control, and range, of which ECRAMs are just one example[24].

The titanium-carbon core layers in $Ti_3C_2T_x$ are highly conductive, while the oxygen and hydroxyl surface functional groups promote a metal-oxide-like surface, enabling fast and highly reversible surface redox activity in protic electrolytes[25,26]. These structural factors enable fast electronic and ionic transport, owing to 2D nanochannels and nanosheet morphology, and therefore ultra-high-rate pseudocapacitive charge storage in multilayer $Ti_3C_2T_x$ MXene films (~400 F $g^{-1}$ at 10 V $s^{-1}$ in 90 nm films[25] and with the capability to perform at even higher charging rates[27,28] of 100 V $s^{-1}$ to 1,000 V $s^{-1}$) – a key requirement for high-speed switching in ECRAMs. In contrast, other 2D materials, with the exception of graphene[29], have not been able to perform at such extreme charging rates. $Ti_3C_2T_x$ has high thermal and chemical stability: dry $Ti_3C_2T_x$ films are stable for several years in ambient conditions, and their water dispersions are stable up to a year[30]. $Ti_3C_2T_x$ retains its chemical structure when annealed to temperatures as high as 830 °C[31] – another key requirement for future ECRAM integration with underlying Si electronics. Finally, the high electrochemical activity of $Ti_3C_2T_x$ in protic acidic electrolytes[25], enables ECRAM operation with protons, the smallest and fastest diffusing cation. Indeed, polymer-ECRAMs that rely on proton-conducting electrolytes[6,16] operated significantly faster than those using Li-based electrolytes[7–14]. Importantly, Li is incompatible with semiconductor fabrication processes. The unique properties of $Ti_3C_2T_x$ (fast charging, thermal/chemical stability, and switching using protons) allow, in a single device, to match the high performance and speed of polymer-ECRAMs while combining it with the integration compatibility of 2D materials and metal oxides.

**Layer-by-layer self-assembly of 2D multilayers**

$Ti_3C_2T_x$ can form a stable water dispersion as the surface terminations on each flake render them negatively charged[18]. This charge enables various thin-film processing techniques for device fabrication, such as layer-by-layer (LbL) self-assembly[32] or interfacial assembly on a wafer scale[33]. To leverage the inherent charge of $Ti_3C_2T_x$, we used an LbL self-assembly process that we developed previously to form multilayered MXene films[32]. We chose LbL assembly because this versatile technique allows precise control over the number of bilayers, denoted n, where each bilayer consists of only a few/single flakes of coplanar $Ti_3C_2T_x$. These films can be further patterned by standard photolithography[34]. The main building block in our LbL-assembled films consists of negatively charged MXene flakes and a positively charged spacer (counter-ion) molecule (**Fig. 1a**). To fabricate devices, we LbL assembled multilayered MXene films at the ECRAM channel and gate, as confirmed by scanning electron microscopy (SEM) (**Fig. 1b**). We used PVA-$H_2SO_4$ as the solid-state proton-conducting electrolyte to connect the ECRAM channel to gate and to leverage the high electrochemical activity of $Ti_3C_2T_x$ in protic acidic electrolytes[25].

**Role of spacer molecules on ECRAM performance**

We studied ECRAM device performance as a function of the number of bilayers and explored three different spacer molecules, tris(3-aminopropyl)amine (TAPA), tris(2-aminoethyl)amine (TAEA), and polyethylenimine (PEI). The different spacer molecules provided evidence of the importance of MXene interlayer chemistry on the properties of electrochemical resistive switching in multilayered MXene films. Transfer curve measurements show a clear and nearly linear tuning of the MXene/TAPA ECRAM channel conductance versus gate voltage (**Fig. 1c**). ECRAMs using other spacer molecules show similar switching but with a reduced dynamic range (TAEA) or slower speed (PEI) (**Supplementary Fig. S1**). While details of electrochemical resistance switching in MXenes are not understood, the mechanism resembles that of conjugated polymers[16,17], where ions inserted from the electrolyte are compensated either electrostatically or via redox reactions by charges on the polymer backbone, thus affecting the carrier density and conductivity of the polymer film.

The TAPA spacer increases the spacing between the MXene flakes from 1.28 nm in pristine $T_3C_2T_x$ to 1.46 nm in the MXene/TAPA multilayer (**Fig. 1d**), leading to a decreased electronic conductivity [~9 x $10^2$ S $m^{-1}$ for (M/TAPA$_6$)] compared to the pristine MXene film (~3 x $10^4$ S $m^{-1}$), while enhancing ionic access to the redox-active sites on the outer Ti atoms of $T_3C_2T_x$. These two factors are both beneficial for neuromorphic computing resulting in lower power consumption and faster switching, respectively. Using TAPA as a spacer between the MXene flakes, the ECRAMs showed nearly no hysteresis under a high-rate gate voltage sweep of 8 V $s^{-1}$, indicating high-speed resistance switching across an appreciable dynamic range irrespective of the number of bilayers (**Fig. 1c**). In contrast, the dynamic range of ECRAMs based on pristine $Ti_3C_2T_x$ was very limited (**Fig. 1c**). We hypothesize that the large MXene/TAPA ECRAM conductance range originates from changes in the electronic structure and electrochemical behavior of the MXene films due to the incorporation of organic molecules. The role of the interlayer molecule on dictating device characteristics is however more complex than simply tuning interlayer spacing. For example, while the smaller TAEA spacer (*d*-spacing = 1.38 nm, **Supplementary Fig. S2**) resulted in similar performance to TAPA but with a more pronounced hysteresis and reduced dynamic range, devices made using the larger PEI spacer (*d*-spacing = 2.15 nm, **Supplementary Fig. S2**) showed substantial hysteresis even at reduced scan rates (**Supplementary Fig. S1**). While further studies are needed for a more detailed understanding of the switching mechanisms and the role of spacer molecules on the improved performance of LbL-assembled MXene-ECRAMs, these results already demonstrate the versatility of the LbL self-assembly technique as a knob to tune device properties. Indeed, the prominent role of interlayer molecules in the electronic behavior of MXene multilayer films has been demonstrated both theoretically[35] and experimentally[19]. While pristine MXene films are metallic, the presence of spacer molecules in multilayer MXene films can result in semiconducting behavior[19] and consequently a larger dynamic range. These results show that the interactions between inserted protons and specific pillaring molecules play an important role in the switching kinetics and mechanisms, significantly affecting the behavior of the different multilayered MXene films, especially when compared to pristine $T_3C_2T_x$ MXene.

Since TAPA resulted in the best performing devices, we used MXene/TAPA ECRAMs for further neuromorphic function characterization. To ensure accurate operation at low power, ANN accelerators require high resistance synaptic devices that span a sufficient dynamic range to accommodate multiple states while tolerating write noise[4,5]. We selected (MXene/TAPA)$_6$ and (MXene/TAPA)$_{10}$ ECRAMs for further characterization due to their highest (>2x) dynamic range among the MXene/TAPA series (**Fig. 1e**). Note that for neuromorphic computing applications it is desirable to operate the synaptic cells linearly across such a relatively small (~2x) dynamic range since this prevents large currents from saturating neurons (e.g., stuck ON cells in PCM arrays).

**Nearly linear switching at high speed and low energy**

To probe device performance under fast pulsing, we used two switches (write and read select) that simulate access devices in an ANN array and ensure accurate readout of both the channel conductance ($G_{SD}$), and the injected write current ($I_{GD}$) (**Supplementary Fig. S3**). **Fig. 2a** shows successful high-speed programming of the (MXene/TAPA)$_{10}$ device using ±1-V 4-µs pulses, followed by 1-µs write-read delay and 0.1-V 10-µs readout. The resulting resistance switching is nearly linear and spans 50x distinct states (**Fig. 2b**), highlighting the potential of these ECRAMs for parallel ANN acceleration[6]. The amount of injected charge (Δ$Q$) per write needed to switch between these states is independent of $G_{SD}$ and equal to 1.6 nC (**Fig. 2b**), corresponding to a constant switching energy of Δ$E$ = 1.6 nJ per write across the ECRAM dynamic range. The area normalized switching energy for (MXene/TAPA)$_{10}$ ECRAMs is Δ$E$ = 80 fJ µm$^{-2}$, meeting the requirements for low-power neuromorphic computing[36].

To ensure low ANN accelerator latency, synaptic devices must exhibit both sub-microsecond write speed and rapid readout. **Fig. 2c** shows successful programming of the (MXene/TAPA)$_6$ ECRAM using 200 ns write pulses, spanning an appreciable dynamic range of 5x at ±3 V. ECRAMs using (MXene/TAPA)$_{10}$ can also be cycled using 200 ns write pulses but over a smaller 1.3x dynamic range (**Supplementary Fig. S4**). Such fast switching is already attainable in large devices (1,000x20 µm$^2$), highlighting the potential of MXene-ECRAMs to meet the high-speed requirements for neuromorphic computing when scaled down. These devices are already orders of magnitude faster than most three-terminal synaptic memories based on 2D materials and metal oxides, typically operating with millisecond write pulses[7–12,15,37,38] (**Fig. 2d** and **Supplementary Table S1**).

**Rapid readout**

Importantly, the conductance of these ECRAMs can be rapidly read following just 1 μs write-read delay (**Fig. 2a**). This is at least a thousand times shorter than other inorganic ECRAMs made with 2D material and metal-oxide channels (**Fig. 2d**). While some previously reported devices have demonstrated non-volatile switching using sub-microsecond write pulses[10,13,14], slow ion kinetics limits these devices to at least millisecond-long write-read delays or require long read pulses (collectively referred to as settling time) to ensure accurate readout of a non-volatile change in device conductance[7–12,14,15] (**Fig. 2d** and **Supplementary Table S1**), severely limiting the overall speed of the device.

**State retention**

To investigate the volatility of conductance states, we programmed the device to a desired source-gate voltage ($V_{SG}$), and continuously monitored $G_{SD}$ with a floating gate (**Fig. 3a**). We observed a bi-exponential decay yielding time constants of ~50 s and ~250 s, respectively. These estimates represent a lower bound as state retention is limited by device architecture and lack of encapsulation, which we have not optimized here. These figures do not translate directly into functional state retention times, which depend on the bit precision requirements of the ANN application where ECRAMs will be used. We note however that our device retention is already adequate for applications where ANN learning occurs either continuously or the learned ANN weights are subsequently transferred to external memory for storage[5].

**High endurance and low write noise**

If these ECRAMs are to be integrated into ANN accelerators, they must demonstrate stable operation during extensive cycling. We observe stable operation following >$10^8$ write-read events in both (MXene/TAPA)$_6$ (**Fig. 3b**) as well as (MXene/TAPA)$_{10}$ ECRAMs (**Supplementary Fig. S5**). In (MXene/TAPA)$_6$, the median channel conductance drifted by only 1.75 % following >$10^8$ write-read events while the device dynamic range and $\Delta Q$ needed to switch between states remained unaffected (<0.1 % change for both). $10^8$ write-read events represent the lower bound for device endurance, as the ECRAM was still fully operational after the last cycle shown in **Fig. 3b**. Such endurance at the early stages of MXene-ECRAM development is very promising as it already exceeds that of FLASH memories ($10^6$ cycles[39]) and it is competitive with state-of-the-art inorganic resistive switches ($10^9$-$10^{12}$ cycles[36]).

Further statistical analysis of the endurance data (**Figs. 3c-d**), obtained by uniformly sampling 3,000 write-read events, reveals that MXene-ECRAMs also display extremely low write noise $\Delta G^2/\sigma^2 > 100$, where $\Delta G$ is the conductance update per write and $\sigma$ its standard deviation. Such high update accuracy is comparable to the best ECRAMs reported to date[6,10,11,16] and significantly exceeds the typical accuracy observed in ReRAMs[5] and PCMs[2] ($\Delta G^2/\sigma^2 < 1$) that limits their ANN performance[4,5]. Using measured MXene-ECRAM switching statistics (**Figs. 3c-d**) as input to ANN simulations, image classification of 8 x 8-pixel handwritten digits yields close to ideal numerical accuracy (**Fig. 3e**), thanks to the linearity and low noise of our MXene-ECRAMs.

**Temperature stability for on-chip integration**

Having established that our ECRAMs fulfill a large number of requirements for efficient neuromorphic computing, we have taken steps towards investigating their on-chip integration compatibility. We annealed a (MXene/TAPA)$_6$ film in vacuum from room temperature to 400 °C using a gradual 15 °C min$^{-1}$ ramp up (total anneal time ~25 min), this way spanning the range of temperatures relevant for back-end-of-line (BEOL) processing (upper limit ~400 °C), and cycled the resulting ECRAM for >$10^8$ write-read events to inspect its reliability. While the heat-treated ECRAM shows increased channel conductance and some loss of linearity, it displays stable operation throughout the $10^8$ write-read events with comparable cycling characteristics to that of the untreated sample (**Fig. 4**). The median channel conductance of the heat-treated sample drifted by only 1.9 % following >$10^8$ write-read events while the device dynamic range and $\Delta Q$ per write remained nearly unaffected (<1 % change for both). The device was still fully operational after the last cycle shown in **Fig. 4**. While significant challenges remain to demonstrate actual BEOL integration, this result marks a fundamental first step towards meeting the temperature requirements for integration into ANN accelerators.

**Device scaling**

Finally, since smaller devices require less $\Delta Q$ per write, the ECRAM write speed, voltage, and switching energy can be further improved by device downscaling, as we have shown previously[6,16]: write duration and switching energy scale linearly with the ECRAM channel area[6,16,17]. For example, while the (MXene/TAPA)$_6$ ECRAM with a 1,000x20 µm$^2$ channel can be programmed across 5x dynamic range using ±3 V 200 ns pulses (**Fig. 2c**), we estimate that a 20x20 µm$^2$ channel (20x reduction in $\Delta Q$) can be programmed across the same range using 20 ns pulses at a low voltage of ±1.5 V and 7.8 pJ per write. Similarly, the relatively high channel conductance demonstrated in this work (sub-mS to mS), which stems from the wide channel dimensions (1,000x20 µm$^2$), can also be reduced by a combination of scaling and materials choice: we estimate a median conductance of 8 µS for a 20x20 µm$^2$ (MXene/TAPA)$_6$ channel, which can be further decreased by exploring other carbides and nitrides from the large family of 2D MXenes[40].

**Conclusions**

This work represents a fundamental first step in realizing the potential of the large family of 2D MXenes as electrochemically tunable materials for high-performance applications, in particular for synaptic memories. To the best of our knowledge, this is the fastest ionic memory based on 2D materials, where the advantages of previously reported materials are combined in a single high-performance device. It also shows that layer-by-layer self-assembly can be used to tackle future challenges by enabling rapid prototyping and fundamental studies of multilayer $Ti_3C_2T_x$ films, MXenes beyond $Ti_3C_2T_x$, and even other 2D materials[41] and 2D heterostructures[42] on a wafer scale[43]. Our work opens several new interesting research avenues, such as elucidating the conductance switching mechanism of 2D materials as a function of their redox state, the effect of interlayer spacing on the electronic and electrochemical behavior of multilayer 2D films, and wafer-scale patterning of MXene films to meet the requirements for device integration into neuromorphic circuitry. $Ti_3C_2T_x$ is just one of the many compositions in the MXene family, where different electronic and electrochemical properties can be dialed in[40]. The high speed and overall performance we demonstrated in ECRAMs here illustrates that MXenes are a new family of electronic materials, ideally suited to devices that operate at the nexus of electrochemistry and electronics.

**Materials and methods**

*MXene preparation.* $Ti_3C_2T_x$ MXene was synthesized according the previously described modified MILD method[44] from the in-house made $Ti_3AlC_2$ MAX phase powder. Briefly, $Ti_3AlC_2$ was made by dry mixing TiC (99.5%, Alfa Aesar), Ti (-325 mesh, 99.5%, Alfa Aesar), and Al (-325 mesh, 99.5%, Alfa Aesar) powders in a 2:1:1 ratio using zirconia balls for 18 h, followed by sintering the powder mixture in flowing Ar atmosphere at 1400 °C for 2 h with a 3 °C min$^{-1}$ ramping rate. The obtained MAX phase block was milled and sieved using a 400 mesh to obtain $Ti_3AlC_2$ powder with particle size smaller than ~38 µm. To synthesize delaminated $Ti_3C_2T_x$, the obtained $Ti_3AlC_2$ powder was slowly added to an etchant solution containing mixture of HCl and LiF (20 mL of 9 M HCl (ACS) + 1.6 g of LiF (98.5%, Alfa Aesar) per 1 g of the MAX phase). The etching was carried out at 35 °C for 24 h while the solution was continuously stirred at 550 rpm using a Teflon coated magnetic stirrer bar. Then, the etched powders were washed several times using DI water to remove the residual acid and aluminum salt complexes. The washing was done each time by adding 50 mL of water to each centrifuge vial (4 vials per 1 g etched powders), shaking the vials, and centrifugation at 3500 rpm for 5 min. This was repeated until the pH of supernatant reaches above 5. At this step the supernatant was decanted, and precipitates were redispersed in 150 mL of DI water. The solution was probe sonicated in an ice bath under flowing Ar for 1 h (35% amplitude, 750 W). Finally, the obtained solution was centrifuged again at 3500 rpm for 1 h and the supernatant was collected, referred to as the $Ti_3C_2T_x$ solution.

*MXene-ECRAM fabrication using LbL self-assembly.* MXene/TAPA multilayers were formed on glass substrates with lithographically patterned Au electrodes (ECRAM channel dimensions, L = 20 µm, W = 1000 µm). Prior to LbL self-assembly, the substrates were sonicated in acetone and ethanol for 20 min each, and subsequently boiled in isopropyl alcohol at 270 °C for 20 min. Pre-cleaned substrates were treated with $O_2$ plasma (Optrel GBR, Multi-stop) for 15 min to create a hydrophilic surface which promotes uniform coating. We used adhesive tape to cover the back side of the substrate during dip-coating. The clean substrates were then loaded into a dipping robot (StratoSequence VI, nanoStrata Inc.). For LbL self-assembly, we used an aqueous $Ti_3C_2T_x$ MXene solution (0.5 g L$^{-1}$) and one of the organic molecules (TAPA, TAEA, or PEI) dissolved in water (1 g L$^{-1}$). One MXene/TAPA bilayer was formed in 8 steps using an automated program (StratoSmart v7.0). First, the pre-cleaned and Au-coated substrates were slowly dipped into the TAPA solution and subsequentially spun in a circle for 5 min. After that, the coated substrate was rinsed 3 times (2 min each) with Milli-Q water, leading to the formation of a positively charged surface. Following these 4 steps, the now positively charged substrate was dipped and spun again in the MXene solution for 5 min and was subsequently rinsed 3 times (2 min each) as described above. Following these 8 steps, a (MXene/TAPA)$_1$ film was formed. This process was then repeated to achieve the desired number of bilayers. The LbL assembled films were dried overnight in a vacuum oven at room temperature. The adhesive tape was removed, and the film was cut using a razor blade to create physically separate channel/gate films of the ECRAM device. We used 1 g poly(methyl methacrylate) (PMMA, Sigma, $M_w$ ~996,000 by GPC, crystalline) dispersed in 10 mL Toluene (Sigma, ACS reagent, ≥99.5%) to coat the Au electrodes in order to prevent shorting the device via Au contact with the gel electrolyte. The PVA-$H_2SO_4$ gel electrolyte was then drop-casted on top of and in-between the channel/gate films to complete the ECRAM device.

*MXene multilayer characterization using SEM and XRD.* We used LbL-assembled MXene multilayer films on Si substrates. Cross-sectional SEM images were collected by field emission SEM (Hitachi S4800, Hitachi Corp., Japan). XRD was performed in air at room temperature. The corresponding XRD patterns were recorded with a PANalytical X'Pert PRO diffraction system using CuKα radiation ($\lambda$ = 1.5418 Å) in the 2θ range from 4° to 20°.

*PVA-H$_2$SO$_4$ gel electrolyte synthesis.* 10 mL Milli-Q water was bubbled through Ar for 1 h and mixed with 1 g of PVA (Sigma, M$_w$ 89,000-98,000, 99+% hydrolyzed). The resulting solution was stirred at 85 °C for at least 2 h until it became transparent, and was subsequently cooled to room temperature. 3 g of H$_2$SO$_4$ (Sigma, >97.5%) was then slowly added to the above solution, followed by stirring for at least 1 hour at room temperature.

*MXene-ECRAM transfer characteristics.* Transfer characteristics were measured in ambient using either a semiconductor parament analyzer (Keithley 4200A-SCS) or a source-meter unit (Keithley 2600) that was controlled using custom LabVIEW code.

*MXene-ECRAM pulsed measurements.* Time-resolved ECRAM cycling experiments were performed in ambient using two waveform generators (33520B and 33522B, 30 MHz, Keysight Technologies) and an oscilloscope (DSOS054A, 500 MHz, Keysight Technologies) that were controlled using custom LabVIEW code. The measurement circuit was assembled using off-the-shelf components on a custom-designed printed circuit board and enclosure. The collected data were analyzed using custom MATLAB code. See **Supplementary Fig. S3** for a measurement schematic and a more detailed explanation.

*MXene-ECRAM retention measurements.* To estimate state retention in ambient, the ECRAM device was first programmed to a desired source to gate voltage ($V_{SG}$), for example $V_{SG}$ = -0.4 V, followed by leaving the gate floating, and continuously measuring $I_{SD}$ using a source-meter unit (Keithley 2600) that was controlled using custom LabVIEW code.

*ANN simulations.* **Fig. 3e** simulations were performed using the crossbar simulator "CrossSim" (Sandia National Laboratory, USA)[4,11], which accounts for experimentally measured ECRAM write noise and switching nonlinearity/asymmetry. Measured cumulative distribution functions (CDF) (**Figs. 3c-d**) were used to generate a lookup table for synaptic weight updates in ANN simulations. Negative weights were obtained by subtracting a bias from the center of the experimental ECRAM conductance range. An ANN with a size of 64x36x10 was trained using a set of 5620 handwritten digits (8 x 8-pixel version of the MNIST handwritten digit dataset[45]).

*MXene/TAPA multilayer anneal*. The (M/TAPA)$_6$ film in **Fig. 4** was annealed in a vacuum probe station (LTMP-4, MMR Technologies) under 2 × 10$^{-4}$ mbar vacuum.

## Acknowledgements

A.M. and A.S. acknowledge financial support from the Semiconductor Research Corporation (SRC), IMPACT nCore Center, Task no. 2966.012. M.K. and M.H. acknowledge financial support by the ÅForsk Foundation (18-461). W.T. and M.H. acknowledge financial support by the Swedish Energy Agency (Energimyndigheten 48489-1). Authors acknowledge Alessandro Enrico at KTH for help with laser patterning, Tyler James Quill at Stanford University for help with ANN simulations, and Toni Moore and Alex Blagojebic for their help at University of Connecticut with MAX phase synthesis. A.V.M. acknowledges Sina Shahbazmohamadi for providing access to chemical laboratories at innovation partnership building (IPB) of the University of Connecticut for part of this project.

## Author contributions

A.M. and M.H. conceived the original idea. A.M. led the project, characterized ECRAM performance, and wrote the manuscript draft. M.K. performed LbL self-assembly and measured SEM. M.K. and A.M. fabricated ECRAMs and measured their transfer curves. W.T. developed the LbL self-assembly of MXene/TAPA multilayers. W.T. and M.K. measured XRD. A.V.M. prepared the MAX phase powder and MXene solutions, provided guidance for MXene film preparation to M.K. and A.M., and helped A.M. draft the manuscript. All authors contributed to developing ideas and interpreting data, and to the writing and preparation of the final manuscript.

**Competing financial interests**

The authors declare no competing financial interests.

**Data availability**

The data and code that support the findings of this study are available from the corresponding authors upon reasonable request.

**Supplementary Information**

Supplementary Figs. S1-S5
Supplementary Table S1

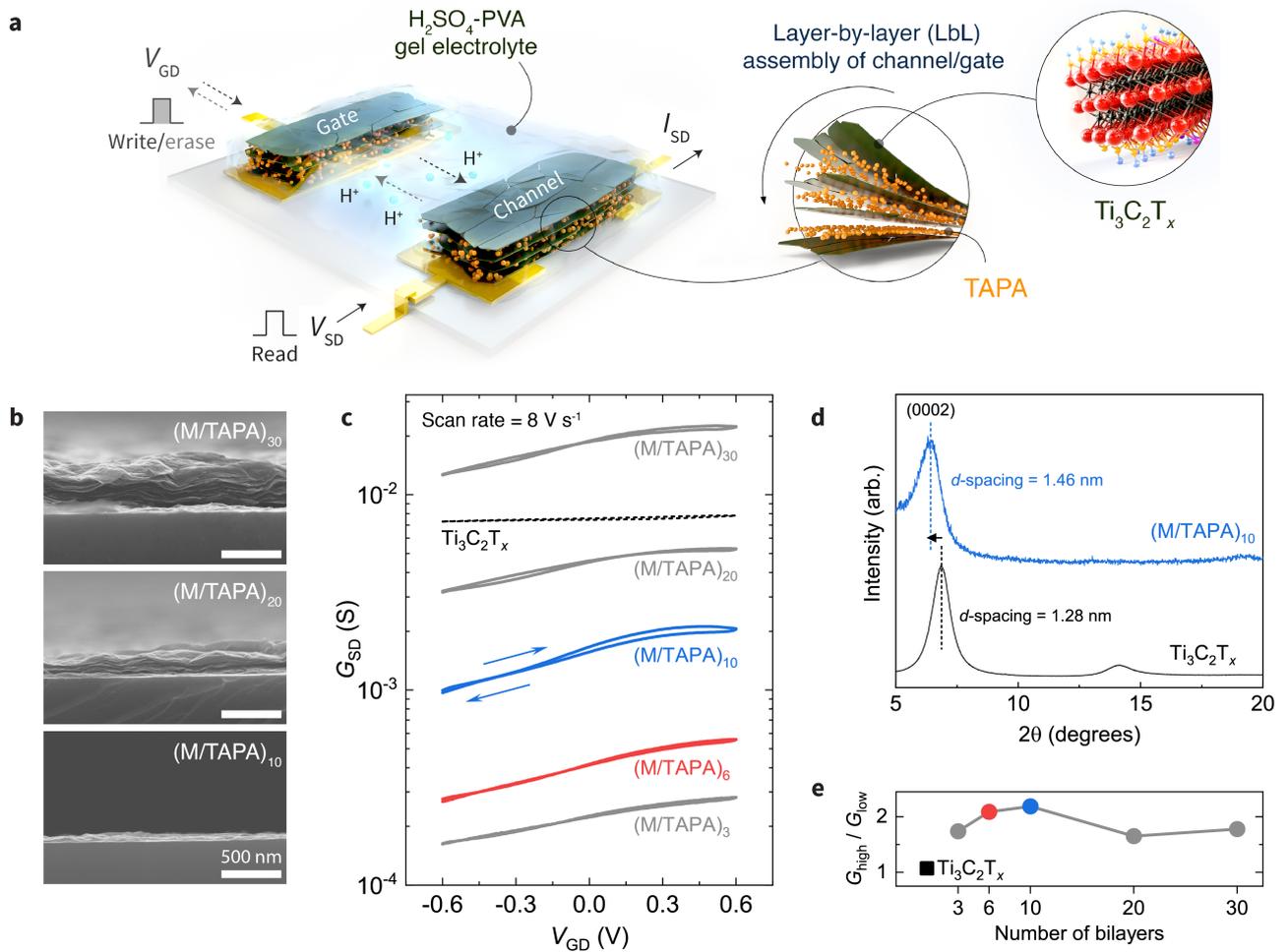

**Figure 1. MXene-ECRAM device architecture and performance optimization using LbL self-assembly.** (**a**) ECRAM device schematic. (**b**) Cross-sectional SEM images of (MXene/TAPA)$_n$ multilayers with n = 10, 20, and 30 bilayers on Si. (**c**) Transfer characteristics of MXene-ECRAMs based on pristine Ti$_3$C$_2$T$_x$ (black dashed) and (MXene/TAPA)$_n$ multilayers (colored), obtained at a high-rate gate voltage sweep of 8 V s$^{-1}$, show nearly linear high-speed resistance switching. (**d**) XRD patterns of pristine Ti$_3$C$_2$T$_x$ (black) and LbL-assembled (MXene/TAPA)$_{10}$ multilayer (blue). Vertical dashed lines mark the (0002) peak positions. (**e**) ECRAM dynamic range versus number of bilayers. (M/TAPA) label abbreviates (MXene/TAPA).

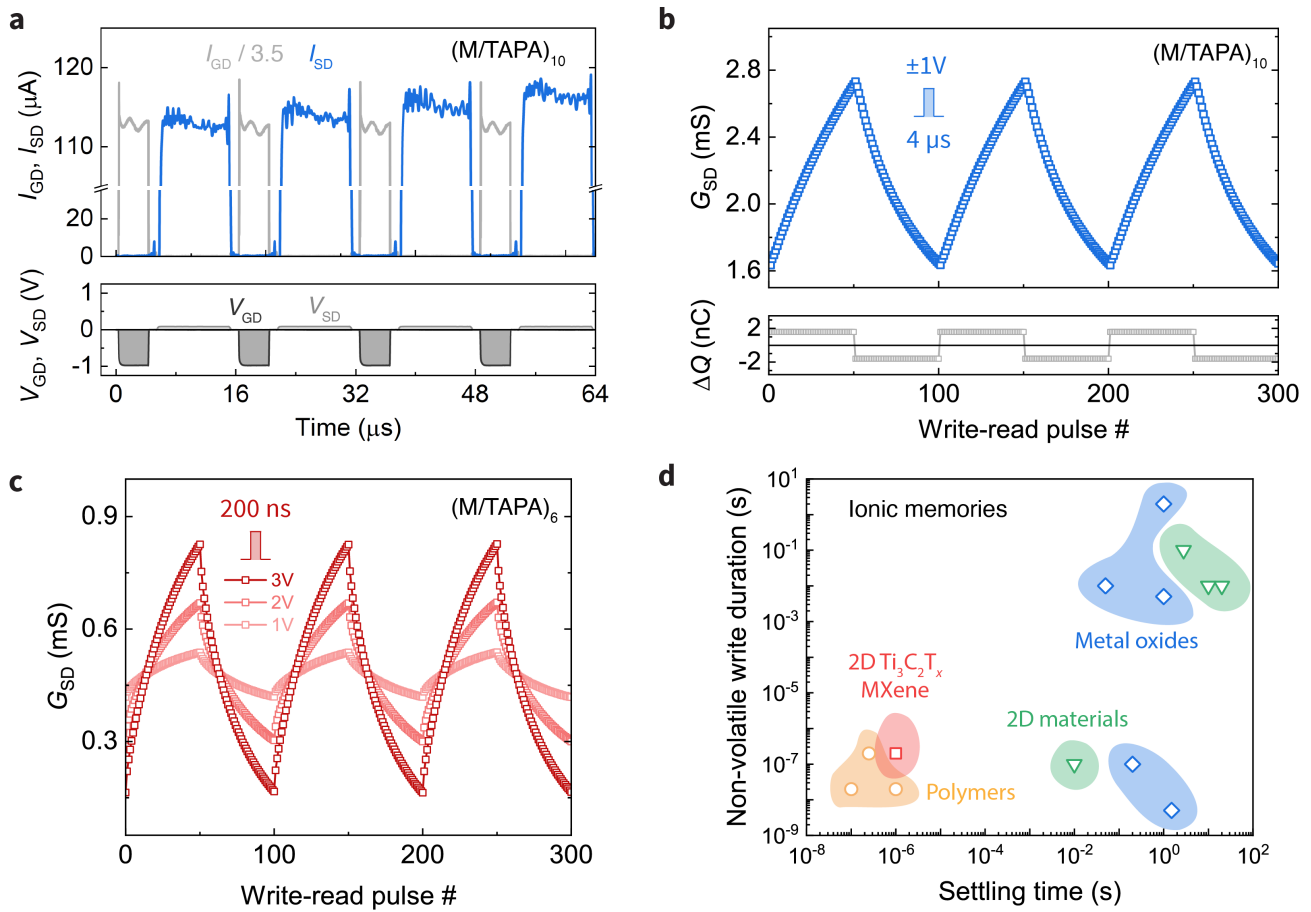

**Figure 2. Fast cycling of MXene-ECRAMs and speed comparison to other ionic memories.** (**a**) Fast pulsed operation of (MXene/TAPA)$_{10}$ ECRAM using ±1-V 4-μs write pulses, followed by 1-μs write-read delay and 0.1-V 10-μs readout. The write current ($I_{GD}$) is divided by 3.5 to fit in the same figure. (**b**) (MXene/TAPA)$_{10}$ ECRAM cycling using ±1-V 4-μs write pulses (top panel) and ΔQ per write (bottom panel). Cycling is nearly linear with constant ΔQ per write across the ECRAM dynamic range. This cycling data was obtained from panel a: channel conductance ($G_{SD}$) was calculated by dividing the measured read current ($I_{SD}$) by the known read voltage ($V_{SD}$), while ΔQ per write was obtained by integrating the measured $I_{GD}$. (**c**) (MXene/TAPA)$_6$ ECRAM cycling using 200 ns write pulses with varying write voltage. Dynamic range increases at higher voltages due to an increase in ΔQ per write. (**d**) Non-volatile write duration versus settling time in ECRAMs using various materials classes: polymers (orange circles), 2D materials (green triangles), metal oxides (blue diamonds), and 2D Ti$_3$C$_2$T$_x$ MXene used in this work (red square). MXene-ECRAMs enable high-speed switching and fast readout comparable to polymer-ECRAMs, and are orders of magnitude faster compared to previously reported ionic memories using 2D materials and metal oxides. (M/TAPA) label abbreviates (MXene/TAPA).

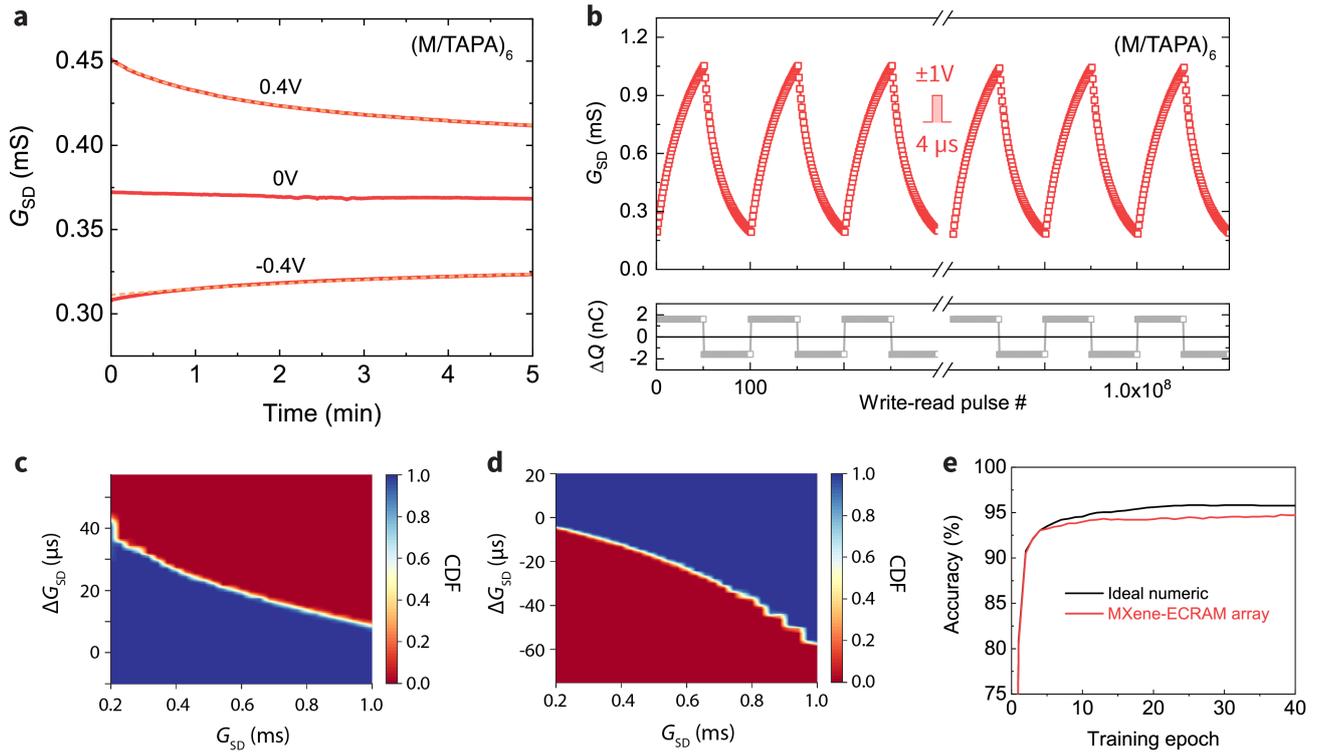

**Figure 3. MXene-ECRAM retention, endurance to >$10^8$ write-read events, and image recognition simulations using experimentally measured ECRAM write noise.** (**a**) (MXene/TAPA)$_6$ ECRAM state retention in ambient (red) after the device has been programmed to the indicated source to gate voltage ($V_{SG}$) and fits to a bi-exponential decay (orange dashed). (**b**) (MXene/TAPA)$_6$ ECRAM endurance to >$10^8$ write-read events (top panel) with stable $\Delta Q$ per write throughout (bottom panel), recorded using ±1-V 4-µs write pulses. (**c-d**) (MXene/TAPA)$_6$ ECRAM $\Delta G_{SD}$ vs $G_{SD}$ switching statistics for write noise characterization for potentiation (**c**) and depression (**d**), obtained by uniformly sampling 3,000 write-read events from panel b. The cumulative distribution function (CDF) represents the probability that $\Delta G_{SD}$ per write is less or equal to the plotted $\Delta G_{SD}$. The small spread around the mean $\Delta G_{SD}$ indicates low write noise ($\Delta G^2/\sigma^2 > 100$). (**e**) Simulated ANN accuracy for 8 x 8-pixel handwritten digit image classification using MXene-ECRAM switching statistics (red) from panels c-d, compared to ideal numerical accuracy (black). (M/TAPA) label abbreviates (MXene/TAPA).

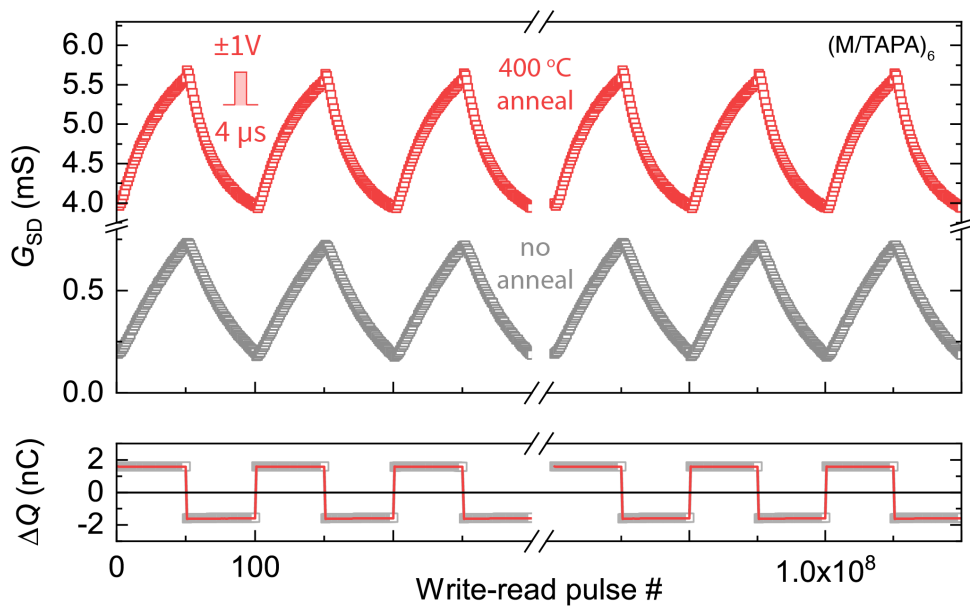

**Figure 4. MXene-ECRAM resilience to high temperature.** Endurance to >$10^8$ write-read events (top panel) with stable Δ$Q$ per write throughout (bottom panel) of MXene-ECRAMs using a (MXene/TAPA)$_6$ film that was annealed up to 400 °C in vacuum (red) versus a non-annealed (MXene/TAPA)$_6$ reference (grey), recorded using ±1-V 4-μs write pulses at room temperature. The (MXene/TAPA)$_6$ film was annealed without the electrolyte. (M/TAPA) label abbreviates (MXene/TAPA).

# Supplementary Information

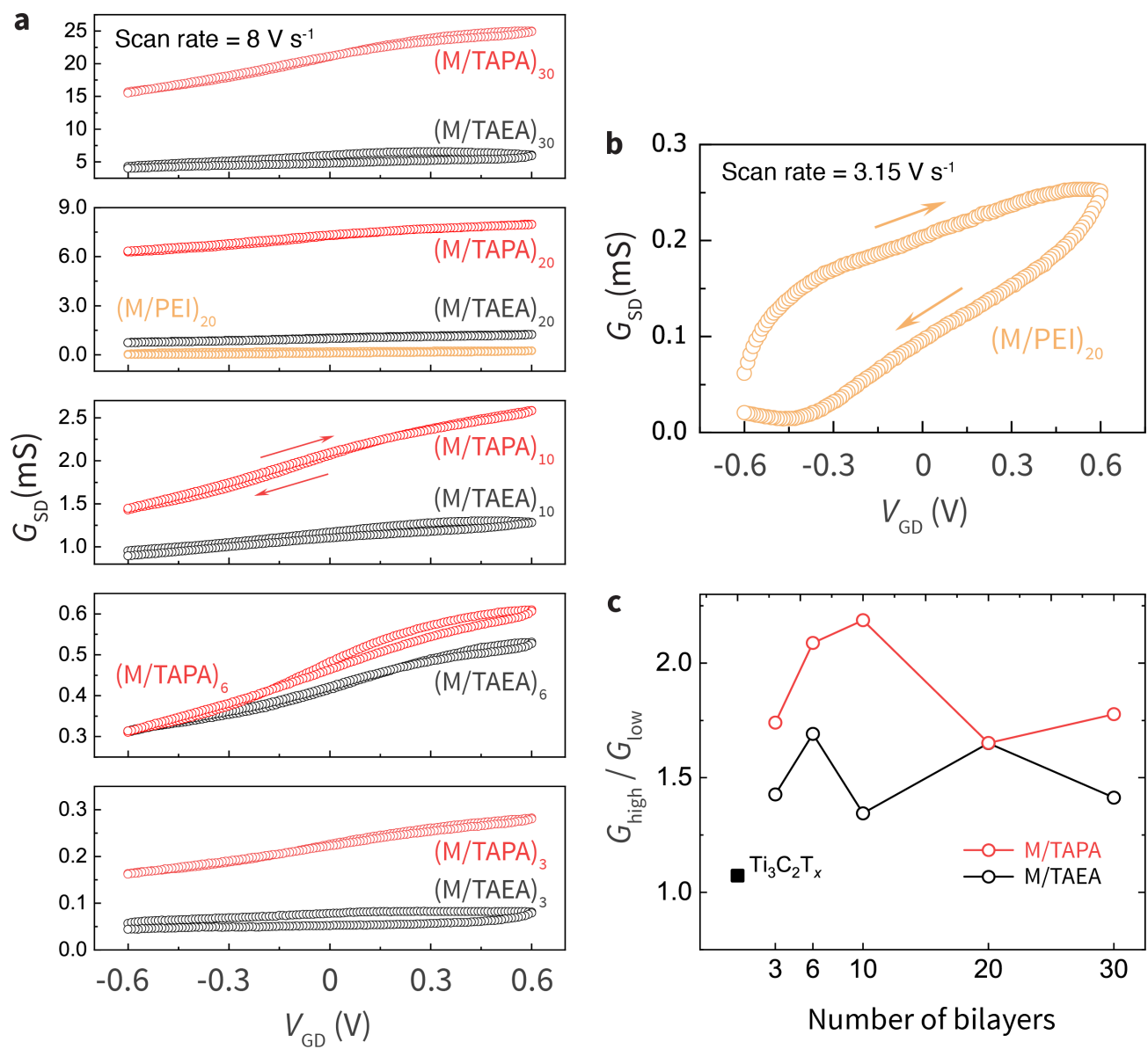

**Supplementary Figure S1. MXene-ECRAM performance versus spacer molecule.** (**a**) Transfer characteristics of MXene-ECRAMs based on (MXene/TAPA)$_n$ (red), (MXene/TAEA)$_n$ (black), and (MXene/PEI)$_n$ versus the number of bilayers, recorded at a scan rate of 8 V s$^{-1}$. MXene/PEI devices were only tested using 20 bilayers due to their slow response (large hysteresis), shown in panel b. (MXene/TAPA)$_n$ devices showed the best performance: nearly no hysteresis and a larger dynamic range compared to (MXene/TAEA)$_n$ irrespective of the number bilayers, shown in panel c. (**b**) Transfer characteristic of (MXene/PEI)$_{20}$ ECRAM shows a pronounced hysteresis even at a reduced scan rate of 3.15 V s$^{-1}$, indicating a slow device response. (**c**) MXene-ECRAM dynamic range versus number of bilayers for (MXene/TAPA)$_n$ and (MXene/TAEA)$_n$ (red and black circles, respectively). The black square shows the limited dynamic range of pristine Ti$_3$C$_2$T$_x$. (MXene/PEI)$_{20}$ dynamic range is 4x but is not shown due to the slow response of the (MXene/PEI)$_{20}$ device, shown in panel b. (M/TAPA), (M/TAEA), and (M/PEI) labels abbreviate (MXene/TAPA), (MXene/TAEA), and (MXene/PEI), respectively.

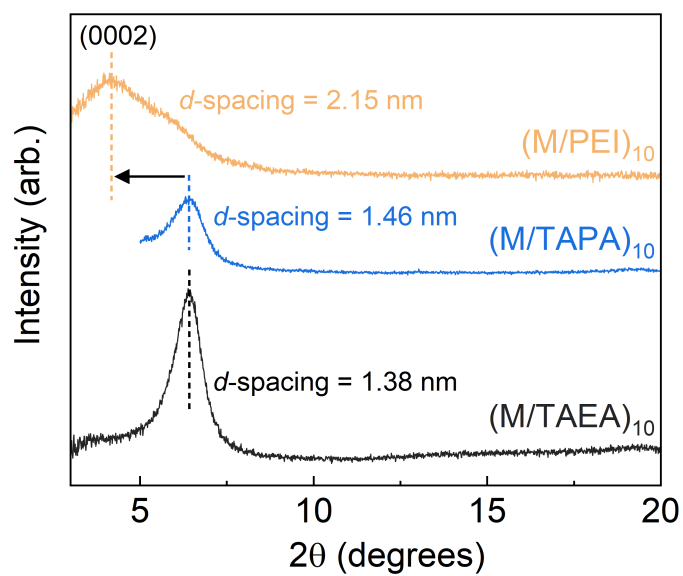

**Supplementary Figure S2. XRD patterns of MXene multilayers using different spacer molecules.** (M/TAEA)$_{10}$ (black), (M/TAPA)$_{10}$ (blue), and (M/PEI)$_{10}$ (orange). Vertical dashed lines mark the (0002) peak positions. (M/TAPA), (M/TAEA), and (M/PEI) labels abbreviate (MXene/TAPA), (MXene/TAEA), and (MXene/PEI), respectively.

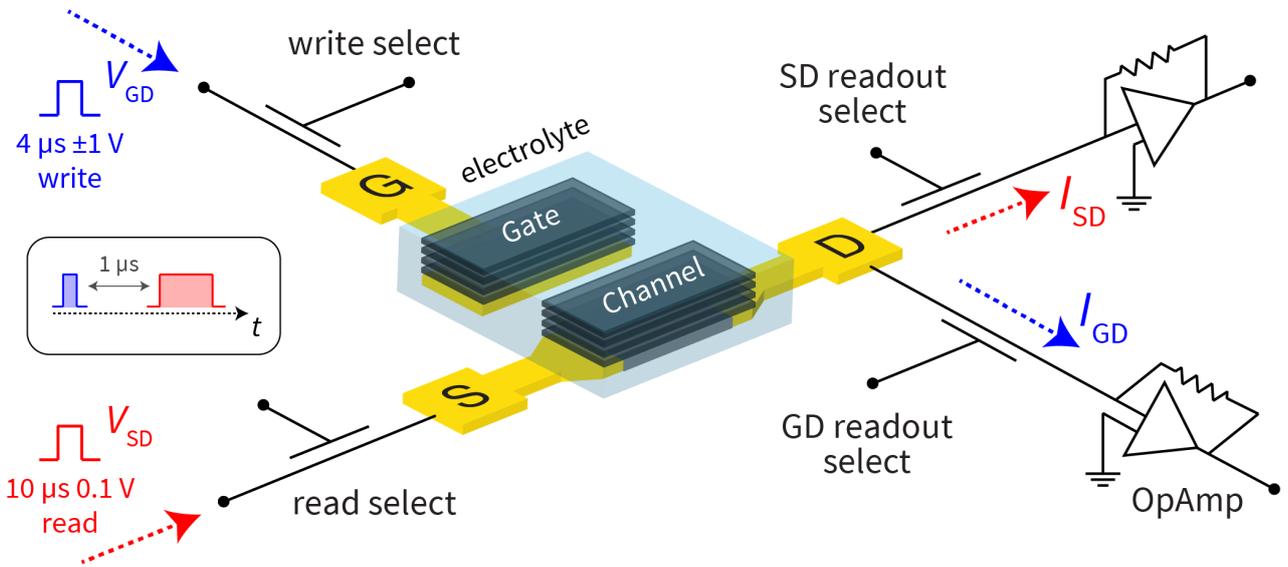

**Supplementary Figure S3. MXene-ECRAM pulsed measurement schematic.** We use a high-speed switch at the ECRAM gate as the write select and a high-speed switch at the ECRAM source as the read select. These switches are temporally synced with the corresponding write (blue) and read (red) pulses, and are turned ON (allowing current flow) only for the duration of the write and read pulses. The above schematic shows the pulsing conditions used to cycle the (MXene/TAPA)$_{10}$ ECRAM in Figs. 2a-b of the main text. The write select is ON for the 4 µs duration of the write pulse and is turned OFF (no current flow) immediately after. The read pulse is delayed by 1 µs, see inset (black rectangle). Following the 1 µs write-read delay, the read select is turned ON for the 10 µs duration of the read pulse and turned OFF immediately after. The ECRAM gate to drain current ($I_{GD}$), i.e. the write current, and the source to drain current ($I_{SD}$), i.e. the read current, are measured using operational amplifiers (OpAmps). The measured OpAmp voltage is converted to current by using known OpAmp gain. We use two additional switches: gate to drain (GD) readout select and source to drain (SD) readout select to ensure accurate readout of $I_{GD}$ and $I_{SD}$, respectively. The GD readout select is ON (allowing current flow) for the duration of the write pulse and is turned OFF (no current flow) immediately after. The SD readout select is synced with the read pulse in similar fashion: the SD readout select is ON (allowing current flow) for the duration of the read pulse and is turned OFF (no current flow) immediately after. The channel conductance ($G_{SD}$) is calculated by dividing the measured $I_{SD}$ by the known read voltage ($V_{SD}$). The write current ($I_{GD}$) is integrated for the duration of the write pulse to obtain the amount of injected charge ($\Delta Q$) per write.

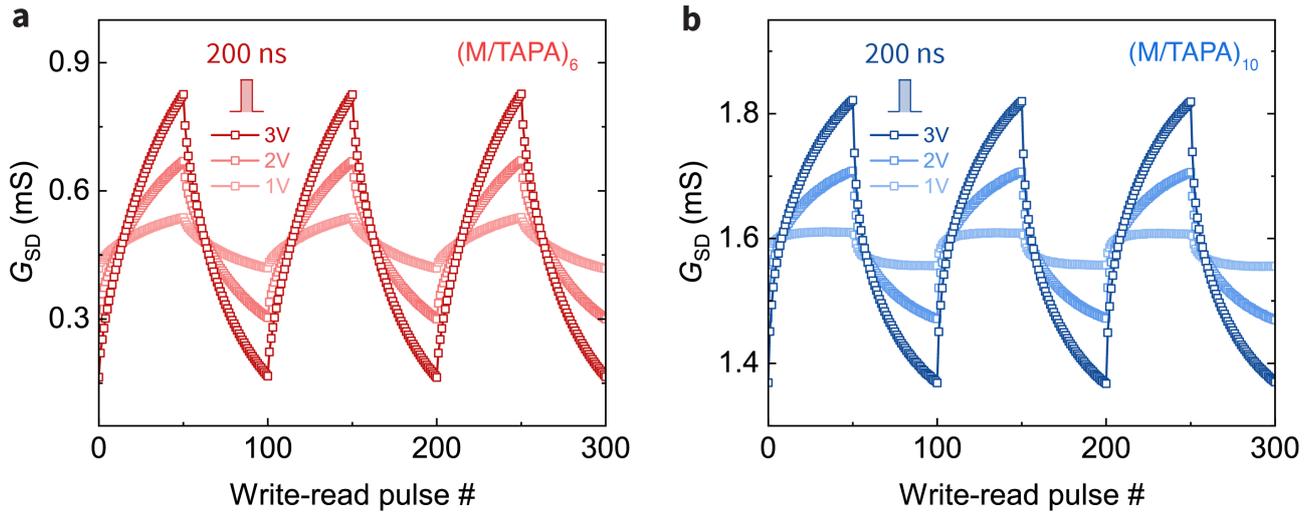

**Supplementary Figure S4. (MXene/TAPA)$_6$ and (MXene/TAPA)$_{10}$ ECRAM cycling using 200 ns write.** Comparison between (MXene/TAPA)$_6$ (**a**) and (MXene/TAPA)$_{10}$ (**b**) ECRAM cycling characteristics using 200 ns write pulses and varying write voltage. In both cases the dynamic range increases at higher voltages due to an increase in $\Delta Q$ per write. (MXene/TAPA)$_6$ dynamic range is significantly larger (5x at ±3 V) compared to (MXene/TAPA)$_{10}$ (1.3x at ±3 V). (M/TAPA) label abbreviates (MXene/TAPA).

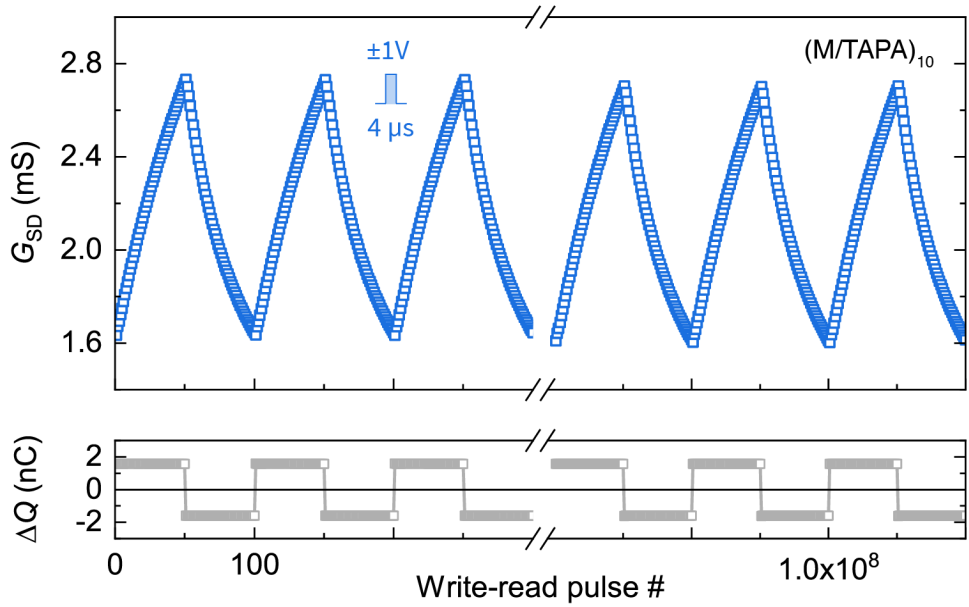

**Supplementary Figure S5. (MXene/TAPA)$_{10}$ ECRAM endurance** to >10$^8$ write-read events (top panel) with stable $\Delta Q$ per write throughout (bottom panel), recorded using ±1-V 4-µs write pulses. Drift = 1.4 %, dynamic range and $\Delta Q$ needed to switch between states remained nearly unaffected (<0.14 % change for both).

**Supplementary Table S1. Performance comparison of various synaptic devices.** MXene- and polymer-based ECRAMs stand out due to their linear high-speed switching at low energy, low noise, and high endurance. Other 3-terminal synaptic memories and ECRAMs based on 2D materials and metal oxides generally suffer from slow write and readout speeds. 2-terminal MXene-based threshold-type resistive switches have been demonstrated[20–23], but lack non-volatility and predictable switching characteristics. 2-terminal multilayered 2D h-BN and 2D $MoS_{2-x}O_x$ devices have shown promising performance, but face the same challenges as conventional 2-terminal memories, such as ReRAM and PCM, namely poor linearity and high write noise. More in-depth analysis of synaptic memory technologies and metrics can be found in recent reviews[4,5,37,38].

| Device | Material | Electrolyte or gate | Non-volatile write duration | Write-read delay or settling time | Write amplitude | Switching energy or power | Endurance | Linearity | # of states | Write noise | Ref. |
|---|---|---|---|---|---|---|---|---|---|---|---|
| 3 terminal ECRAM | **2D MXene** Multilayer $Ti_3C_2T_x$/TAPA | $H_2SO_4$-PVA | 200 ns | 1 μs | ±1 V | 80 fJ μm$^{-2}$ | >10$^8$ writes | Med/High | >50 | Low | This work |
| | **Polymers** | | | | | | | | | | |
| | p(g2T-TT) | EMIM:TFSI PVDF-HFP | 20 ns | 100 ns | ±1 V | <10 fJ μm$^{-2}$ | >10$^8$ writes | Excellent | >100 | Low | 16 |
| | PEDOT:PSS | EIM:TFSI PVDF-HFP | 20 ns | 1 μs | ±1 V | <10 fJ μm$^{-2}$ | >10$^8$ writes | High | >100 | Low | 16 |
| | PEDOT:PSS | Nafion | 200 ns | 250 ns | ±1 V | 4 fJ μm$^{-2}$ | >10$^8$ writes | Excellent | >50 | Low | 6,16 |
| | **2D materials** | | | | | | | | | | |
| | Graphene | LiClO$_4$-PEO | 10 ms | ~20 s | ±50 pA | <500 fJ | >500 | High | >250 | - | 7 |
| | WSe$_2$ multilayer | LiClO$_4$-PEO | 100 ms | 2.8 s | +1.2V/-0.4V | >30 fJ | - | Med/High | >50 | - | 8 |
| | α-MoO$_3$ | LiClO$_4$-PEO | 10 ms | 10 s | ±2.5 V | 1.8 pJ | - | Med | >50 | Low | 9 |
| | **Metal oxides** | | | | | | | | | | |
| | Li$_x$TiO$_2$ | LiClO$_4$-PEO | 10 ms at 80 °C | ΔG settles in ~50 ms | ±0.3 V | ~30 fJ μm$^{-2}$ | >10$^6$ writes | High | >250 | Low | 12 |
| | Li$_{1-x}$CoO$_2$ | LiPON | 2 s | ~1 s | ±0.75 mV | ~50 aJ μm$^{-2}$ | - | Med | >200 | Low | 11 |
| | WO$_3$ | LiPON | 5 ns | 1.5 s | ±0.3 mA | 2 pJ | >10$^5$ writes | High | >200 | Low | 13 |
| | WO$_3$ | HfO$_2$ | 100 ns | 0.2 s | ±4 V | - | >10$^7$ writes | High | >40 | - | 14 |
| 3 terminal | **2D materials** | | | | | | | | | | |
| | MoS$_2$/h-BN heterostructure | Gr gate | 1 ms (50 ns) | - | ±12 V (±20 V) | 7.3 fJ | - | Med | >30 | - | 46 |
| | MoS$_2$ monolayer | Si/SiO$_2$ gate | 1 ms | - | ±30 V | - | - | Poor | >2 | - | 47 |
| | Graphene | Pt/TiN/Si/Al$_2$O$_3$ gate | 1 s | - | ±5 V | 5 mJ | >500 | - | >16 | Med | 48 |
| | **Metal oxides** | | | | | | | | | | |
| | α-Nb$_2$O$_5$ | Li$_x$SiO$_2$ gate | 100 ns to 10 ms | 10 ms to 1s | +3.6V/-3.4V | 20 fJ μm$^{-2}$ | >10$^6$ writes | Med/High | >32 | Low | 10 |
| | WO$_3$ | Nafion | 5 ms | ~1 s | ±0.5 μA | ~860 aJ μm$^{-2}$ | >2x10$^4$ writes | Poor | >100 | - | 15 |
| 2 terminal | **2D MXene** | | | | | | | | | | |
| | Multilayer Ti$_3$C$_2$T$_x$ | - | 10 ns set 500 μs reset | - | ±2 V | - | >10$^6$ writes | Poor | >2 | - | 20 |
| | Multilayer Ti$_3$C$_2$T$_x$ | - | 1 ms | - | ±3 V | - | >200 writes | Poor | >2 | - | 21 |
| | Multilayer Ti$_3$C$_2$T$_x$ | - | 50 ns, volatile | - | +4 V | 0.35 pJ | >100 writes | Poor | >2 | - | 22 |
| | Multilayer Ti$_3$C$_2$T$_x$ | - | 360 ms, volatile | - | -4.5 V | 25.9 pJ | - | Poor | >2 | - | 23 |
| | **2D materials** | | | | | | | | | | |
| | Multilayer MoS$_{2-x}$O$_x$ | - | 100 ns | - | +3V/-4V | - | >10$^7$ writes | - | 2 | - | 49 |
| | Multilayer h-BN | - | >20-200 μs, many pulses needed for nonvolatility | - | +0.7V/-0.1V | 600 pW | - | Poor | >2 | - | 50 |
| | Multilayer h-BN | - | 1 ms set | - | 5.8V set | 20 fJ for volatile write | >8x10$^4$ cycles | Poor | >25 | - | 51 |
| | Multilayer Li$_x$MoS$_2$ | - | 1 ms | - | ±4 V | - | >10$^4$ writes | Poor | ~100 | High | 52 |
| | **Example ReRAM** TiO$_x$/HfO$_x$ | - | ~10 ns | - | ~1.2 V | 1-24 pJ | ~10$^9$-10$^{13}$ writes | Poor | 2-100 | High | 53 |
| | **Example PCM** Ge$_2$Sb$_2$Te$_5$ | - | ~50 ns | - | 5 V | 2-50 pJ | ~10$^9$-10$^{12}$ writes | Poor | 2-100 | High | 53 |